\documentclass[a4paper]{revtex4}
\usepackage{graphicx}
\usepackage{fancyhdr}
\usepackage{amsmath}
\usepackage{soul}
\newcommand{\TeV}{{\ensuremath\rm TeV}}
\newcommand{\GeV}{{\ensuremath\rm GeV}}
\newcommand{\MeV}{{\ensuremath\rm MeV}}

\newcommand{\pb}{{\ensuremath\rm pb}}
\newcommand{\eqn}{equation}

\newcommand{\lb}{\left(}
\newcommand{\rb}{\right)}

\newcommand{\lam}{\lambda}

\def\D0{\slash\!\!\!\!\!\!\!\!\!\:D0}
\newcommand{\HS}{\texttt{HiggsSignals-1.3.1}}
\newcommand{\HB}{\texttt{HiggsBounds-4.2.0}}

\newcommand{\oblique}{Altarelli:1990zd,Peskin:1990zt,Peskin:1991sw,Maksymyk:1993zm}

\newcommand{\fr}{\frac}

\usepackage{color}

\begin{document}


%

%
%

\title{ Constraining {the} Inert Doublet Model   \vspace{0.5cm}}
\vspace*{1.0truecm}
\author{Agnieszka Ilnicka\vspace{0.2cm}}
\email{ailnicka@physik.uzh.ch}
\affiliation{Institute of Physics, University of Zurich, Winterthurstrasse 190, CH-8057 Zurich\vspace{0.2cm}}
\affiliation{Physics Department, ETH Zurich, Otto-Stern-Weg 1, CH-8093 Zurich\vspace{0.2cm}}
\author{Maria Krawczyk\vspace{0.2cm}}
\email{krawczyk@fuw.edu.pl}
\affiliation{Faculty of Physics, University of Warsaw, ul. Pasteura 5, 02- 093 Warsaw, Poland \vspace{0.2cm}}
\author{Tania Robens}
\email{Tania.Robens@tu-dresden.de}
\affiliation{TU Dresden, Institut f\"ur Kern- und Teilchenphysik, Zellescher Weg 19, D-01069 Dresden, Germany\vspace{0.5cm}}

\begin{abstract}
We give a survey on current constraints on the Inert Doublet Model parameter space, including all theoretical as well as experimental limits from collider and astrophysical data. For allowed regions in the parameter space, we provide total production cross sections for the pair-production of scalars at the 13 \TeV~ LHC and propose benchmarks scenarios which should be investigated by the LHC experiments at Run II.
\end{abstract}

\maketitle

\thispagestyle{fancy}


\section{Introduction}
 One of the simplest models for a scalar dark matter is the Inert Doublet Model (IDM), a version of a Two Higgs Doublet Model with an exact $Z_2$ symmetry \cite{Deshpande:1977rw}.  It contains one SU(2) doublet of spin-0 fields, playing the same role as the corresponding  doublet in the SM, {  with the SM-like Higgs particle. This doublet is even under the $Z_2$ symmetry}, while the second doublet, being $Z_2$-odd,   {is not involved in the mass generation and does not interact with fermions}. This so called inert or dark doublet  contains 4 scalars, two charged and two neutral ones, with the lightest neutral scalar being a natural DM candidate. {Due to the $Z_2$ symmetry, all particles from the dark doublet are pair-produced.}

{This model was studied in order to provide a heavy Higgs boson \cite{Barbieri:2006dq} as well as a lighter  Higgs boson, to be produced at LHC \cite{Cao:2007rm}.  It was considered as a model with a \lq\lq{}perfect example\rq\rq{}  of {a} WIMP \cite{LopezHonorez:2006gr,Honorez:2010re,Dolle:2009fn,Sokolowska:2011aa}.  It leads to an interesting pattern of the Universe evolution, towards the Inert phase as given by the IDM, with one, two or three
phase transitions \cite{Ginzburg:2010wa}. Furthermore, the IDM can provide a strong first-order phase transition \cite{Hambye:2007vf,Chowdhury:2011ga,Borah:2012pu,Gil:2012ya, Blinov:2015vma},  which  is one of the Sakharov conditions needed to generate a baryon asymmetry of the Universe.  
After a discovery in 2012 of a SM-like Higgs particle, many analyses have beed performed for the IDM, using  Higgs {collider} data, as well as astrophysical measurements, {see eg. {\cite{Swiezewska:2012eh,Gustafsson:2012aj,Arhrib:2013ela,Krawczyk:2013jta}}}. {In addition,} proposals {were made} how to search for  dark scalars at {the} LHC {in the di- or multilepton channel} \cite{Dolle:2009ft,Swaczyna,Gustafsson:2012aj}.

Recently, also the important issue of vacuum (meta)stability in the IDM has been discussed, and it was found that additional (possibly heavy)  scalars can have a strong impact on it \cite{Kadastik:2011aa,Goudelis:2013uca,Swiezewska:2015paa,Khan:2015ipa} \footnote{Similar solutions can be found in a simple singlet extension of the SM Higgs sector, cf. e.g. \cite{Pruna:2013bma, Robens:2015gla} and references therein.}.  Especially on the eve of the LHC run II, the determination of the regions of parameters space which survive all current theoretical and experimental constraints is indispensible. Therefore, we here provide a survey of the IDM parameter space, after all constraints are taken into account. We found that the main sources of these are run I LHC Higgs data as well as dark matter relic density (Planck) and direct dark matter search at LUX.

\section{The Inert Doublet Model}


The $D$-symmetric  potential of the IDM has the following form:
\begin{equation}\begin{array}{c}
V=-\fr{1}{2}\left[m_{11}^2(\phi_S^\dagger\phi_S)\!+\! m_{22}^2(\phi_D^\dagger\phi_D)\right]+
\fr{\lambda_1}{2}(\phi_S^\dagger\phi_S)^2\! 
+\!\fr{\lambda_2}{2}(\phi_D^\dagger\phi_D)^2\\[2mm]+\!\lambda_3(\phi_S^\dagger\phi_S)(\phi_D^\dagger\phi_D)\!
\!+\!\lambda_4(\phi_S^\dagger\phi_D)(\phi_D^\dagger\phi_S) +\fr{\lambda_5}{2}\left[(\phi_S^\dagger\phi_D)^2\!
+\!(\phi_D^\dagger\phi_S)^2\right],
\end{array}\label{pot}\end{equation}
with all parameters {taken as} real
(see e.g. \cite{Ginzburg:2010wa}). 
The vacuum state in the IDM is given by \footnote{In a 2HDM with the potential $V$ {given by eqn.} (\ref{pot}) different vacua can exist, e.g. a mixed one with $\langle\phi_S\rangle\neq0$, $\langle\phi_D\rangle\neq0$,  an inertlike vacuum with $\langle\phi_S\rangle=0$, $\langle\phi_D\rangle\neq0$, or {{even a charge breaking vacuum}} see~\cite{Ginzburg:2010wa,Sokolowska:2011aa,Sokolowska:2011sb,Sokolowska:2011yi}. }:
\begin{equation}
\langle\phi_S\rangle =\frac{1}{\sqrt{2}} \begin{pmatrix}0\\ v\end{pmatrix}\,,\qquad \langle\phi_D\rangle = 
\begin{pmatrix} 0 \\ 0  \end{pmatrix}, \label{dekomp_pol}
\end{equation}
{where $v\,=\,246\,\GeV$ denotes the vacuum expectation value (vev).}
After minimization,  {with $v$ fixed,} the model has in total 6 free parameters. We here choose 
\begin{\eqn}\label{eq:freepars}
M_h,\,M_H,\,M_A,\,M_{H^\pm}, \lam_2,\,\lam_{345}
\end{\eqn}
as independent input parameters, with $\lam_{345}\,=\,\lam_3+\lam_4+\lam_5$. 
\section{Theoretical and experimental constraints}\label{sec:constraints}
We have performed an extensive scan on the parameter space of the IDM, taking all current theoretical as well as experimental constraints into account. We give a short description of these in the following paragraphs and refer to \cite{us} for more details.
 Several of the observables below have been calculated using the public tools 2HDMC \cite{Eriksson:2009ws} and micrOMEGAs \cite{MO2013}.
\paragraph{Theoretical constraints}
Our model is subject to several theoretical constraints, which we briefly list here. First, vacuum stability as well as positivity pose several constraints on relations of the couplings \cite{Nie:1998yn, Swiezewska:2012ej}
\begin{\eqn}
\lam_1\,\geq\,0,\,\lam_2\,\geq\,0,\,\lam_3+\sqrt{\lam_1 \lam_2}>0,\,\lam_{345}+\sqrt{\lam_1 \lam_2}\,>0,\, \frac{m_{11}^2}{\sqrt{\lam_1}}\,\geq\,\frac{m_{22}^2}{\sqrt{\lam_2}},
\end{\eqn}
where the last condition guarantees that the inert vacuum ~\ref{dekomp_pol} is global. We furthermore require the scattering matrix to be unitary, leading to a limit of $|\L_i|\,\leq\,8\,\pi$ on the eigenvalues of specific hypercharge/ isospin scattering matrices. Finally, we require all couplings to have maximal absolute values of $4\,\pi$ to stay within the perturbative range. Note that choosing $M_H$ as the dark matter candidate leads to
\begin{\eqn}
M_H\,\leq\,M_A,M_{H^\pm}.
\end{\eqn}
\paragraph{Experimental measurements and constraints}
The discovery of a Higgs boson by the LHC experiments and the measurement of its mass leads to fixing $M_h$. Following the ATLAS/ CMS mass combination \cite{Aad:2015zhl}, which renders $M_h\,=\,125.09\,\pm\,0.21\,\pm\,0.11\,\GeV$, we use 
\begin{\eqn}\label{eq:mh}
M_h\,=\,125.1\,\GeV
\end{\eqn} 
throughout our study. Furthermore,  {upper} limits exist on the total width of this particle \cite{Khachatryan:2014iha, Aad:2015xua} of  {$\sim\,22\,\MeV.$}\\
The decay widths of the electroweak gauge bosons have been extremely well measured by the LEP experiments \cite{Agashe:2014kda}. Being conservative, we require
\begin{\eqn}
M_{A,H}+M_H^\pm\,\geq\,m_W,\,M_A+M_H\,\geq\,m_Z,\,2\,M_H^\pm\,\geq\,m_Z.
\end{\eqn} 
Constraints from null-searches at colliders as well as agreement with the 125 \GeV~ Higgs signal coupling strength an $95\,\%$ confidence level are implemented via the publicly available tools \HB~ \cite{Bechtle:2008jh, Bechtle:2011sb, Bechtle:2013wla} and \HS~ \cite{Bechtle:2013xfa}. Constraints from electroweak precision tests are taken into account via the oblique parameters $S,\,T,\,U$ \cite{\oblique}, which are equally required to agree within $95\,\%$ C.L. taking full correlations between them into account \cite{Baak:2014ora}. We require the charged Higgs to decay within the collider, leading to a lower limit of $\Gamma_\text{tot}(H^\pm)\,\geq\,6.58\,\times\,10^{-18}\,\GeV$. Furthermore, bounds from reinterpreted SUSY searches at LEP render $M_H^\pm\,\geq\,70\,\GeV$ \cite{Pierce:2007ut} and   rule out regions where \cite{EspiritoSanto:2003by}
\begin{\eqn}
M_A\,\leq\,100\,\GeV,\,M_H\,\leq\,80\,\GeV, \Delta M\,\geq\,8\,\GeV.
\end{\eqn}
{With this set of requirements we also stay out of {the} region ruled out by {the} recent reinterpretation of SUSY bounds from LHC \cite{Belanger:2015kga}}.
As discussed above, one of the advantageous features of our model is the fact that it comes with a stable dark matter candidate. Therefore, current limits on dark matter relic density as well as direct detection cross sections need to be respected. We consider the first by requiring 
\begin{\eqn}
\Omega_c\,h^2\,\leq\, 0.1241
\end{\eqn}
to be in agreement with the upper limit on relic density from the last Planck collaboration results \cite{Planck:2015xua}. Limits from direct detection are taken into account {in form of} the last LUX limits \cite{Akerib:2013tjd}.
\section{Results}\label{sec:results}
After fixing $M_h$  (\ref{eq:mh}), we have performed an extensive scan on the remaining 5 parameter space, taking all theoretical and experimental constraints into account.  If not stated otherwise, $\lam_2\,\in\,[0;4.5]$ and $\lam_{345}\,\in\,[-1.5;2]$. Masses were varied from $0\, \lb M_H\rb$ to 1 \TeV~ for $M_H\,\lb M_A,\,M_{H^\pm}\rb$ respectively. The scan has {been performed in} three subsequent steps
\begin{itemize}
\item{}In the first step, we use the mass {and coupling} regions as given above. All points, whether allowed or excluded, are kept; all exclusion criteria for a specific parameter point are memorised. We test all theoretical constraints, as well as the total width of the $125\,\GeV$ Higgs boson {and of} the charged scalar, and electroweak precision observables. Also limits from collider exclusions recast from SUSY searches and Z/W widths are applied {at this stage}.
\item{}Only  points which have passed the above constraints {are then}  checked against limits from the Higgs signal strength and collider searches via \HB~/ \HS~. 
\item{} Points which passed {all} {above}  bounds are then furthermore tested against Planck and LUX data .
\end{itemize}
Constraints on the masses are usually obtained through an interplay of several constraints rendering a direct correlation difficult. However, we found that both couplings $\lam_2$ and $\lam_{345}$ can be directly related to a few constraints. This can e.g. be observed directly from figure \ref{fig:lconstr}, which shows the limits in the $\lb \lam_2,\,\lam_{345} \rb$ plane after scan step 1 and in the $\lb M_H,\,\lam_{345}\rb$ plane after scan step 3, respectively.
{A clear cut from perturbativity on the largest allowed value of $\lam_2$ arises from} the upper bound on the quartic coupling of the H, leading to $\lam_2\,\leq\,\frac{4}{3}\,\pi\,\approx\,4.19$ \cite{Arhrib:2013ela,Swiezewska:2012ej} \footnote{As $\lam_2$ only plays a role in the self-couplings of the dark scalars, it is not a relevant parameter for the remainder of our discussion.}. We furthermore observe a $\lam_2\,\lb \lam_{345}\rb$ dependent positivity induced lower bound on $\lam_{345}\,\lb \lam_2 \rb$. After performing all steps of the scan, we additionally find that for masses $M_H\,\geq\,60\,\GeV$, the strongest constraint on $\lam_{345}$ indeed stems from direct dark matter experiments, leading to $|\lam_{345}|\,\lesssim\,1.2$ for masses $M_H\,\lesssim\,1\,\TeV$. {On the other hand, in the region where the channel $h\,\rightarrow\,H\,H$ is kinematically open, {it was found} that additional strong constraints stem from the SM Higgs signal strength measurements, leading to $|\lam_{345}|\,\lesssim\,0.02$ {(}{see} {also} {e.g. \cite{Krawczyk:2013jta,Krawczyk:2015vka})}. Furthermore, an interplay of limits from weak gauge boson decays and dark matter constraints basically rules out the whole region where $M_H\,\leq\,48\,\GeV$.}
A more dedicated discussion of all other constraints will be given in \cite{us}. As relevant results, we here exemplarily display all {points forbidden by a particular constraint} as well as allowed regions in the $(M_H,\,M_{H^\pm})$ and $(M_H,\,M_A)$ plane in figure \ref{fig:massrats}. We found that the ratios of all dark scalar masses are $\lesssim\,1.5$, implying a relatively strong degeneracy. We found that a blind scan favours high mass regions for all dark scalars \footnote{This holds independently of the numerical cutoff value of the scan range.}.
\begin{center}
\begin{figure}[t]
\begin{minipage}{0.45\textwidth}
\includegraphics[width=\textwidth]{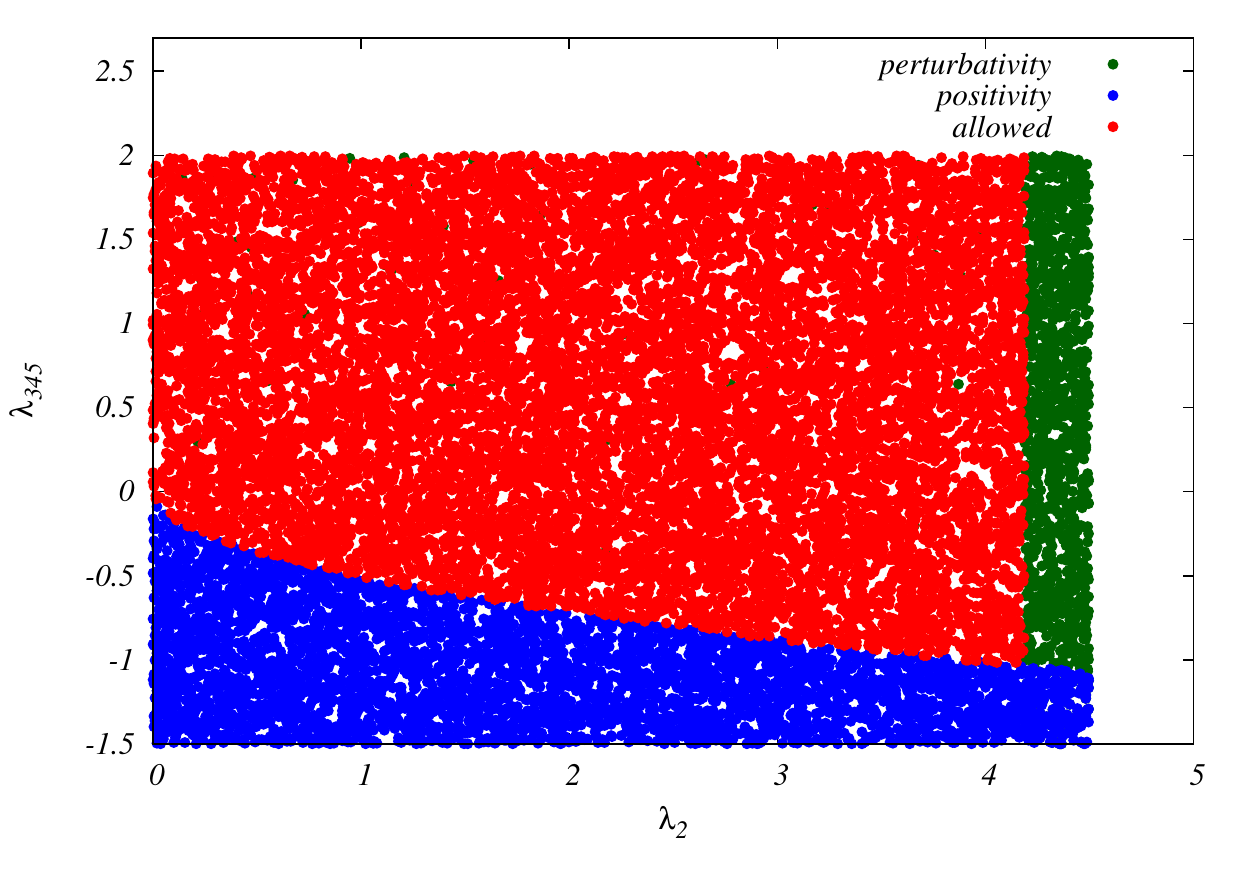}
\end{minipage}
\begin{minipage}{0.45\textwidth}
\includegraphics[width=\textwidth]{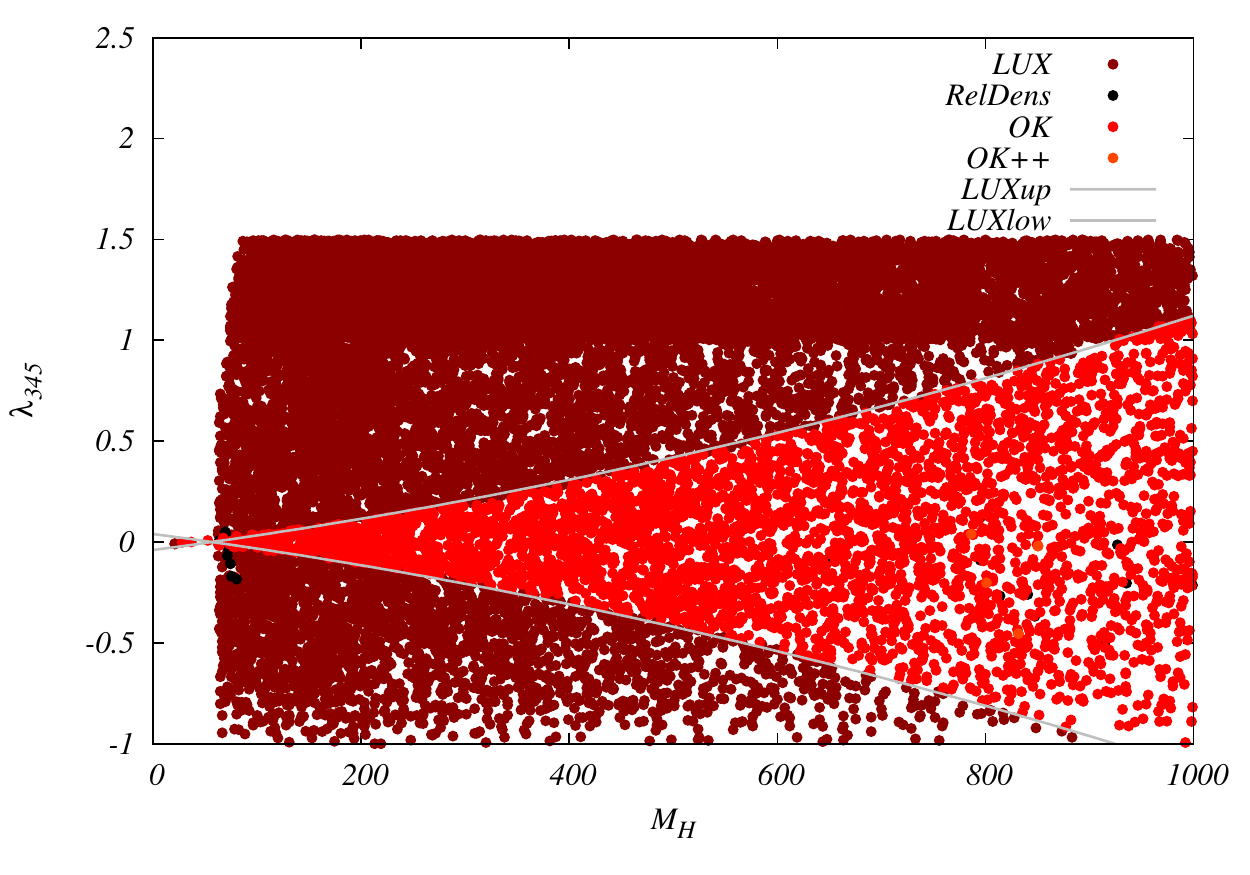}
\end{minipage}
\caption
{\label{fig:lconstr} Left: {Constraints} from positivity {(forbidden points in blue, lower left curve)} and perturbativity {(forbidden points in green, right stripe)} in the $(\lam_2,\,\lam_{345})$ plane. {Points which survive perturbativity and positivity constraints after scan step 1 are displayed in red.} Right: {Step 3 scan:} allowed (red) and forbidden regions in the $(M_H,\,\lam_{345})$ plane from {the direct DM detection  {from} LUX (dark red) and relic density {from} Planck measurements} (black).} 
\end{figure}
\end{center}
\begin{center}
\begin{figure}
\begin{center} 
\begin{minipage}{0.45\textwidth}
\includegraphics[width=\textwidth]{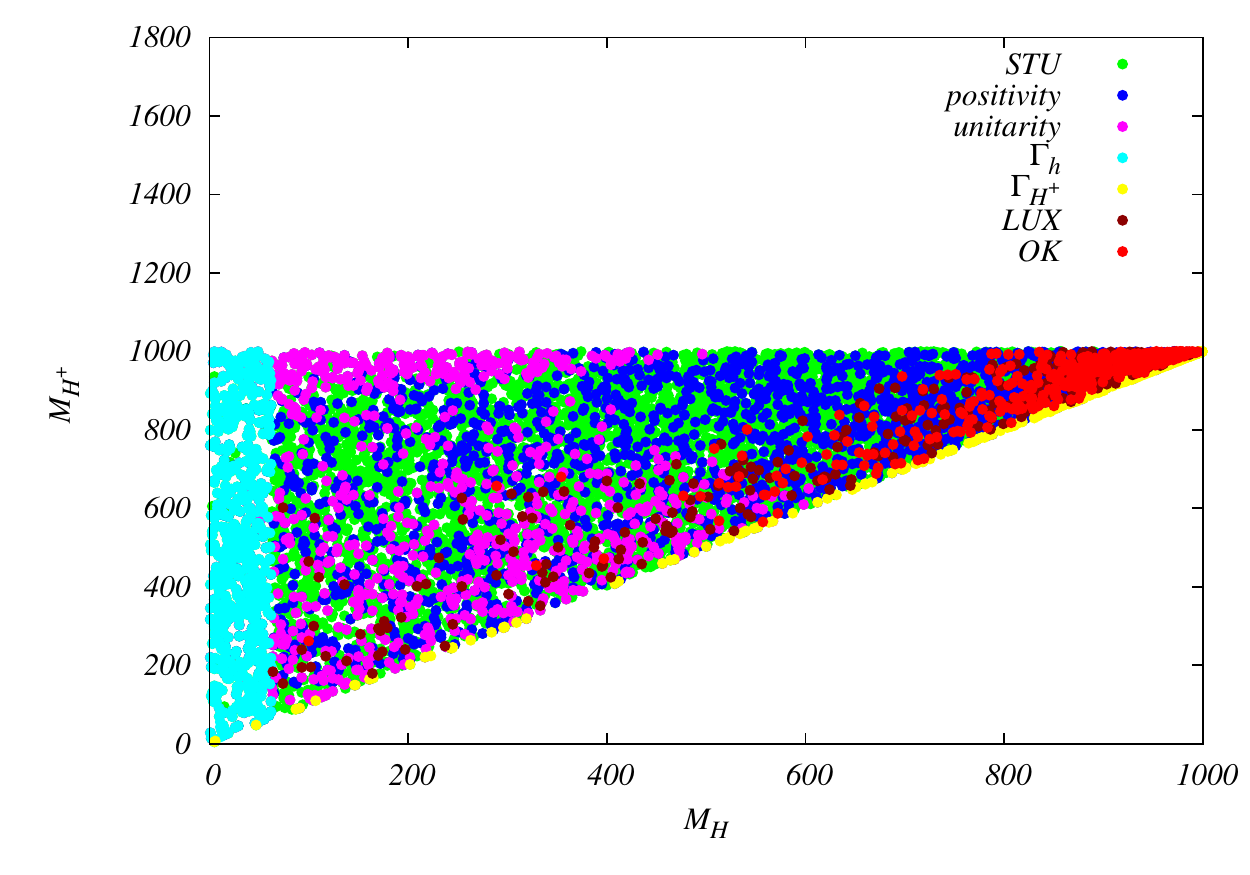}
\end{minipage}
\hspace{1 cm}
\begin{minipage}{0.45\textwidth}
\includegraphics[width=\textwidth]{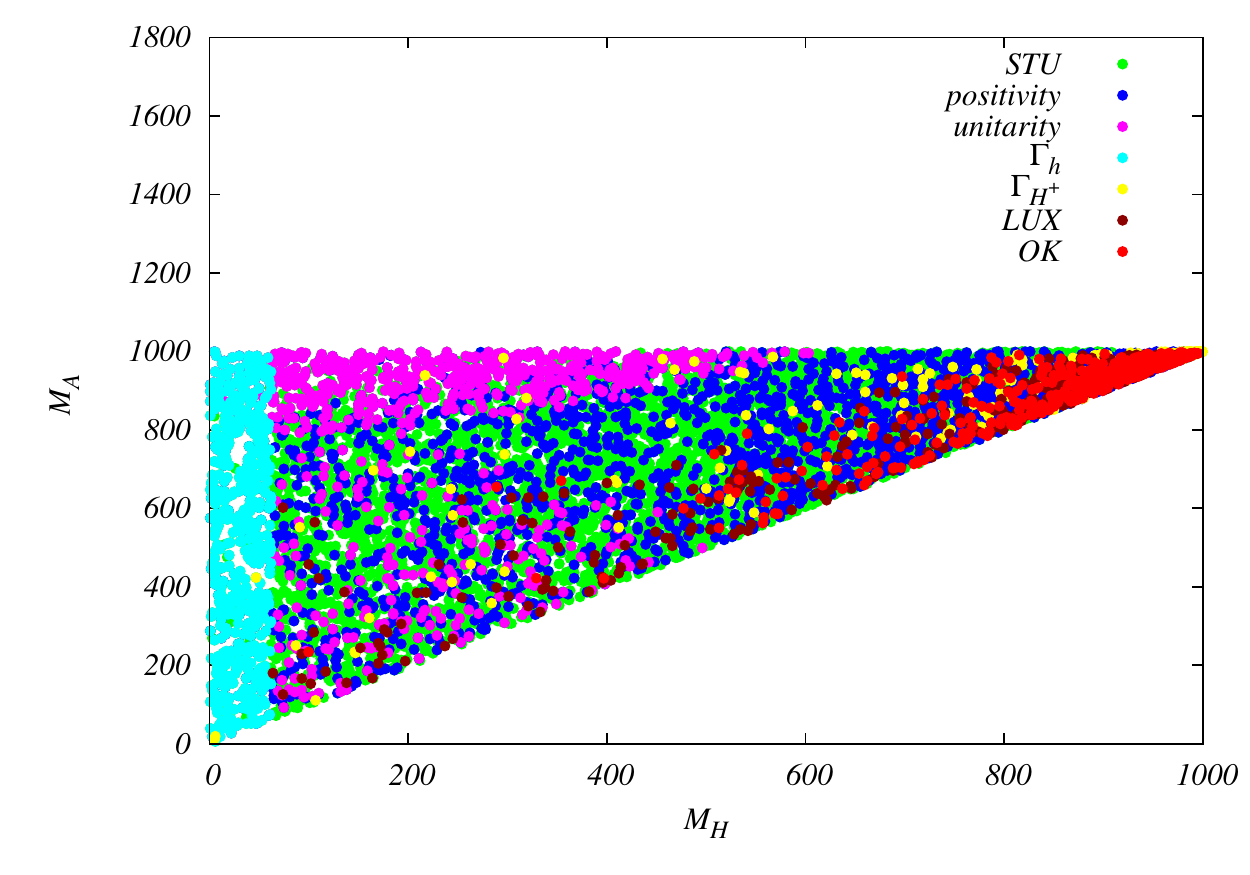}
\end{minipage}
\end{center}
\caption{\label{fig:massrats} $M_H$ vs $M_{H^\pm}$ {\sl (left)} and $M_H$ vs $M_A$ {\sl (right)} after all constraints have been taken into account. {Color coding refers to the constraints which rule out parameter points, where the scan follows the steps as described in section \ref{sec:results}. Points labelled "OK" have survived all theoretical and experimental bounds.}}
\end{figure}
\end{center}

\section{Benchmarks}
Out of all allowed regions in parameter space, we then chose 5 benchmark points which should be tested by the LHC experiments at the current LHC run \cite{usbm}. We give parameter specifications and total cross sections {for $H\,A$ and $H^+\,H^-$ pair-production}, {which were obtained using Madgraph5 \cite{Alwall:2011uj} together with an UFO IDM input file \cite{Degrande:2011ua,Goudelis:2013uca}}  (cf. also figure \ref{fig:xsec}).

\begin{itemize}
\item {\bf Benchmark I: low scalar mass; $H A: 0.371 (4) \pb,\,H^+\,H^-: 0.097 (1) \pb$}
\begin{\eqn*}
M_H=57.5\,\GeV,\,M_A=113.0\,\GeV,M_{H^\pm}= 123\,\GeV,\,\lam_2\,\in[0;4.2],\,\lam_{345}\,\in[-0.015;0.015] 
\end{\eqn*}
\item {\bf Benchmark II: low scalar mass; $H A: 0.226 (2) \pb, H^+ H^-: 0.0605 (9) \pb$}
\begin{\eqn*}
M_H=85.5\,\GeV,\,M_A=111.0\,\GeV,M_{H^\pm}= 140,\,\GeV,\,\lam_2\,\in[0;4.2],\,\lam_{345}\,\in[-0.015;0.015] 
\end{\eqn*}
\item {\bf Benchmark III: intermediate scalar mass; $H\,A: 0.0765 (7)\pb$, $H^+\,H^-: 0.0259 (3) \pb$}
\begin{\eqn*}
M_H=128.0\,\GeV,\,M_A=134.0,\GeV,M_{H^\pm}= 176.0\,\GeV,\,\lam_2\,\in[0;4.2],\,\lam_{345}\,\in[-0.05;0.05] 
\end{\eqn*}
\item {\bf Benchmark IV: high scalar mass, mass degeneracy; $H,A: 0.00122(1) \pb$, $H^+ H^-:  0.00124 (1) \pb$}
\begin{\eqn*}
M_H=363.0\,\GeV,M_A= 374.0\,\GeV,M_{H^\pm}= 374.0\,\GeV,\,\lam_2\,\in[0;4.2],\,\lam_{345}\,\in[-0.25;0.25] 
\end{\eqn*}
\item {\bf Benchmark V: high scalar mass, no mass degeneracy; $H,A: 0.00129 (1) \pb$, $H^+ H^-: 0.000553 (7) \pb$}
\begin{\eqn*}
M_H=311.0\,\GeV,M_A= 415.0\,\GeV, M_{H^\pm}\,=\,447.0\,\GeV,\lam_2\,\in[0;4.2],\,\lam_{345}\,\in[-0.19;0.19] 
\end{\eqn*}

\end{itemize}
While benchmarks I and II are exceptional points in a sense that the allowed parameter space is extremely constrained in the low mass region, benchmarks III to V are more typical, as these parts of the parameter space are more highly populated. Furthermore, for scenario IV the production cross sections for $HA$ and $H^+ H^-$ have a similar order of magnitude.
\begin{figure}
\begin{minipage}{0.45\textwidth}
\includegraphics[width=\textwidth]{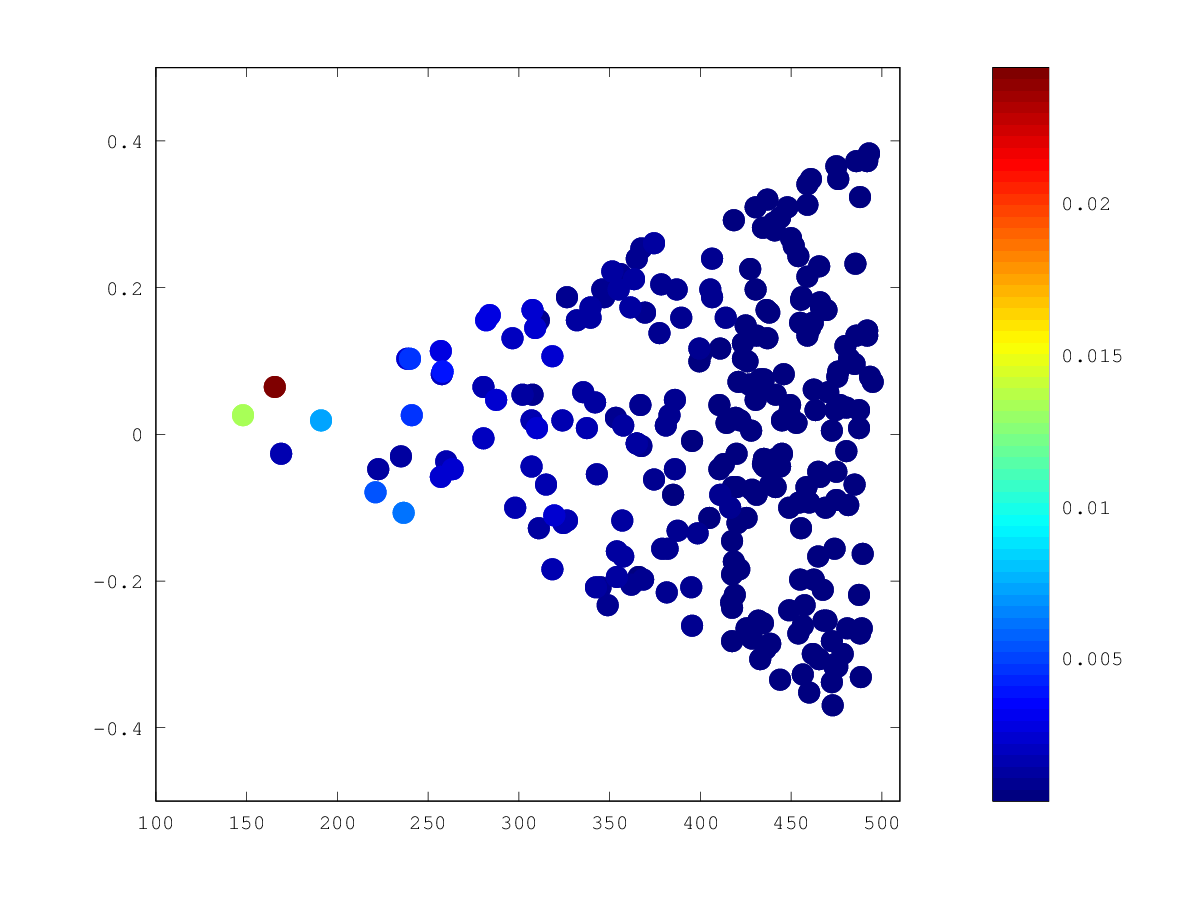}
\end{minipage}
\begin{minipage}{0.45\textwidth}
\includegraphics[width=\textwidth]{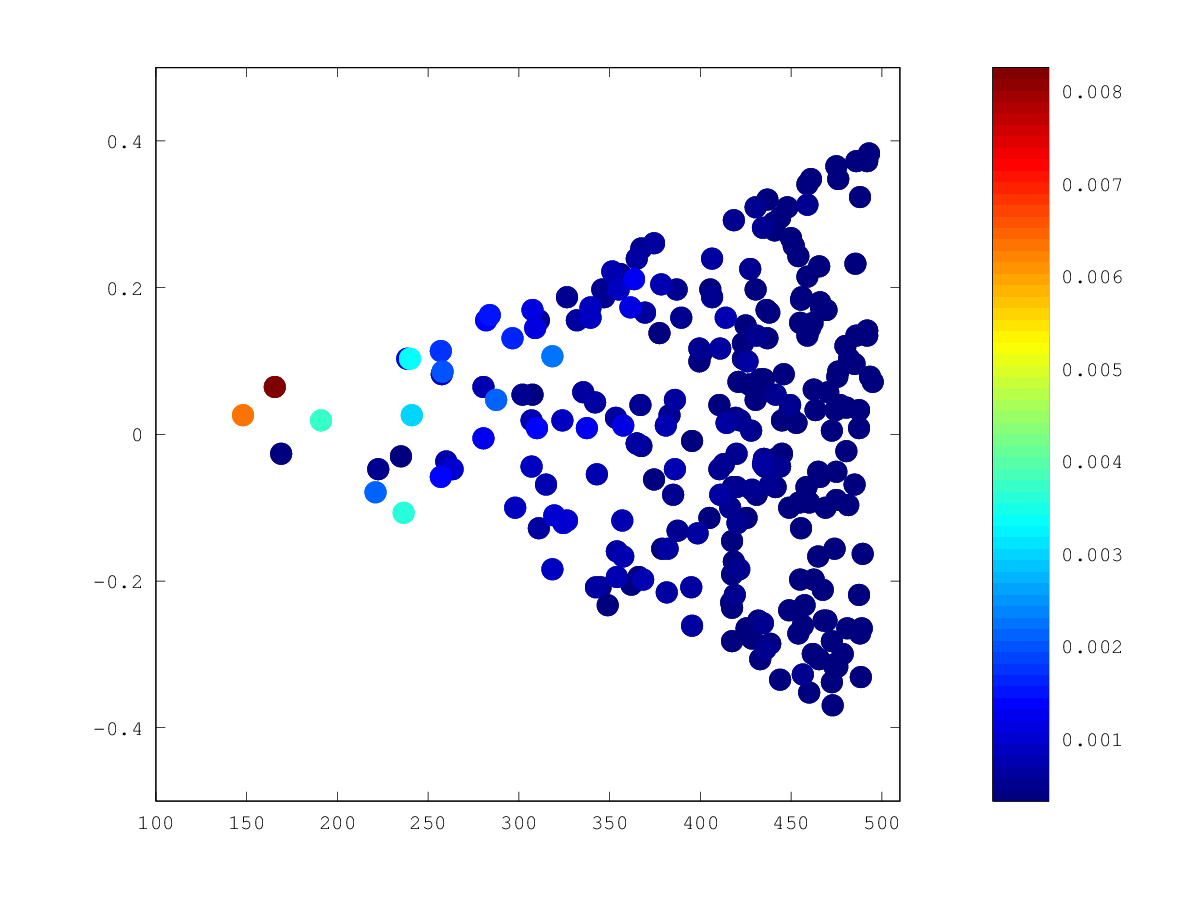}
\end{minipage}
\caption{\label{fig:xsec} Production cross sections in \pb~ at a 13 \TeV~ LHC for $H A$ {\sl (left)} and $H^+\,H^-$ {\sl (right)}, in the $\lb M_H,\,\lam_{345} \rb$ plane. {Points presented here are allowed by all constraints mentioned above.}}
\end{figure}
\section{Conclusions}
The Inert Doublet Model provides a viable extension of the Standard Model scalar sector. While allowing for the EWSB as in the SM, it additionally introduces a dark scalar {sector} {with a good} dark matter candidate. We here discussed all currently available constraints on this model and identified the most stringent constraints on {dark matter couplings} $\lam_2$ and $\lam_{345}$.  {We found that scalar masses are still allowed in basically the whole mass range considered here, with   a lower limit of $M_H\,\geq\,48\,\GeV$, arising }due to an interplay of LEP results and dark matter constraints. {Generically,} the parameter space opens up for higher {dark scalar masses}. {For $M_H\,\gtrsim\,300\,\GeV$,} mass degeneracies are observed, with mass ratios below $1.5$. We also provide pair production cross-sections for 5 benchmark scenarios, which could be investigated experimentally {at} LHC run {II}. 







\begin{acknowledgments}
MK thanks {the} organizers for an excellent organization and scienific atmosphere at the Toyama workshop and for financial support. She acknowledges discussions with  I. Ivanov and  M. Sampaio. {TR wants to thank  A. Goudelis, B. Herrmann, O. Stal and T. Stefaniak for useful discussions and R. Frederix for MG5 support.}  {We also thank B. Swiezewska and D. Sokolowska for discussion on the IDM as well as the Higgs Cross Section working group for encouraging us to investigate possible benchmarks.} {This research was supported by the DAAD grant PPP Poland Project 56269947 "Dark Matter at Colliders" } and by the Polish grant NCN OPUS 2012/05/B/ST2/03306
(2012-2016). {The work of AI is supported by the 7th Framework Programme of the European Commission through the Initial Training Network HiggsTools PITN-GA-2012-316704.} 
\end{acknowledgments}

\bigskip 
\bibliography{lita}

\end{document}